\documentclass[letterpaper, 10 pt, conference]{ieeeconf}  

\usepackage{graphicx}
\usepackage{url}
\usepackage{listings}
\usepackage{xcolor}

\definecolor{maroon}{rgb}{0.5,0,0}
\definecolor{darkgreen}{rgb}{0,0.5,0}
\lstdefinelanguage{XML}
{
  basicstyle=\ttfamily\footnotesize,
  morestring=[s]{"}{"},
  morecomment=[s]{?}{?},
  morecomment=[s]{!--}{--},
  commentstyle=\color{darkgreen},
  moredelim=[s][\color{black}]{>}{<},
  moredelim=[s][\color{red}]{\ }{=},
  stringstyle=\color{blue},
  identifierstyle=\color{maroon}
}

\IEEEoverridecommandlockouts                              

\title{\LARGE \bf
Crowdsourced bi-directional disaster reporting and alerting on smartphones in {Lao PDR}
}

\author{Lutz Frommberger$^{1}$ and Falko Schmid$^{1}$
\thanks{$^{1}$Lutz Frommberger and Falko Schmid are with the International Lab for Local Capacity Building (Capacity Lab) at the Faculty for Mathematics and Informatics at the University of Bremen, Germany. Capacity Lab, Enrique-Schmid-Str. 5, 28359 Bremen, Germany. Email: {\tt\small\{lutz,schmid\}@capacitylab.org}, WWW: {\tt\small www.capacitylab.org}}}

\begin{document}

\maketitle
\thispagestyle{empty}
\pagestyle{empty}

\begin{abstract}
Natural disasters are a large threat for people especially in developing countries such as Laos. ICT-based disaster management systems aim at supporting disaster warning and response efforts. However, the ability to directly communicate in both directions between  local and administrative level is often not supported, and a tight integration into administrative workflows is missing. In this paper, we present the smartphone-based  disaster and reporting system Mobile4D. It allows for bi-directional communication while being fully involved in administrative processes. We present the system setup and discuss integration into administrative structures in Lao PDR.
\end{abstract}

\section{INTRODUCTION}

Natural disasters are a threat for people in any country in the world. But especially in developing countries. They can have severe consequences that affect people's lives. It is widely recognized that natural disasters are a main reason for poverty as they "reduce or eliminate equal access to opportunities and, therefore, to development" \cite{alcantara2002geomorphology}. Due to climatic change and increasing populations, the effect of natural disasters and the damage caused appears to rise dramatically.

In Lao People's Democratic Republic (Lao PDR), the Mekong river and its confluences are of critical importance for many of the inhabitants, and a large fraction of the population lives near those rivers. Thus, tropical storms and the resulting floods can have a severe impact on the whole country. As an example, the typhoon Ketsana that struck the southern provinces of Lao PDR in late September 2009 resulted in 180,000 people being directly affected and an estimated damage of 58 Million US-\$ \cite{mrc-report2009}. Agricultural sector was hit the hardest, which aggravated rice production and, thus, food security. But significant damage was also done to the transport sector by destroying roads and bridges. Lao government estimated the loss of GDP caused by this single event at 0.4\%, that is, about 20 Million US-\$ \cite{mrc-report2009}.

Large-scale disasters like this are a great challenge for people affected. Equally, they are a great challenge for governmental administrative units (GAUs) which are in charge of disaster response, which is a difficult issue under large-scale disaster conditions. But in developing countries, people  are also often confronted with problems on a smaller scale, e.g., local outbreaks of human, plant, or animal diseases. While initially not having a large impact on the overall country, these smaller incidents can have severe consequences for affected individuals. It may also easily occur that, e.g., diseases can spread and affect others, and, thus, become larger scale problems.

\begin{figure}
\centering
\includegraphics[width=\columnwidth]{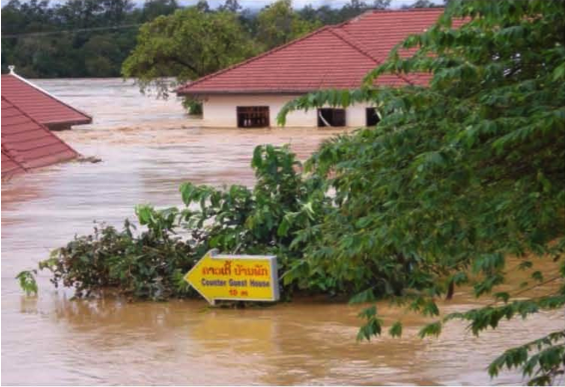}
\caption{Flood in Lao PDR after typhoon Ketsana in September 2009. It directly affected 180,000 people.}
\label{ketsanaflood}
\end{figure}

In any disaster case, the flow of information is a critical issue. This applies both for communication from the local level towards administration, and vice versa. On the one hand, detailed information from affected regions  is essential for the appropriate GAUs to organize disaster response and provide needed support, and on the other hand, information on upcoming disasters or updates on the situation are needed for local people to take the right actions. A seamless flow of information between different administrative levels (e.g., from district to province level) is also essential for efficient disaster response. 

To account for this, we report on Mobile4D, a bi-directional location-based disaster alerting and reporting system based on smartphones that, on the one hand, allows for sending out emergency warnings from the administration to affected people and, on the other hand, to report disasters at the local level as a crowdsourcing effort.

This paper is organized as follows: First, we describe related work on disaster management systems, especially in developing countries. Then we give a detailed description of the Mobile4D system in Sect.~\ref{design}, highlighting the situation in Lao PDR, Goals, and architecture and features of the system. Section~\ref{results} reports on a first test in Lao PDR before the paper closes with a conclusion.

\section{RELATED WORK}

Several ICT frameworks and systems related to disasters are in use, most of them targeting the developed world.  
In developing countries, additional issues have to be faced. As 
 \cite{smarajiva:tsunamilessons:2005} point out,
effective warning systems require "not only the use of ICTs, but also the existence of institutions that allow for the effective mobilization of their potential", so the effective inclusion of administrative units play a critical role.

Sahana \cite{currion2007open} is a complex modular Open Source disaster management toolkit targeting at large-scale disasters, especially for organizing and coordinating disaster response. It has been successfully applied in many lesser developed countries.
A review on Geohazard Warning Systems is given in \cite{bhattacharya2012review}.

Mobile devices gain increasing importance in disaster cases. Disaster alert systems based on SMS show a good impact in developing countries \cite{Mahmud:SMSDisaster2012}.
\cite{fajardo2010mobile} present a Android smartphone based disaster alerting system which focusses mostly on routing issues in the disaster response phase.
In general, the use of smartphones can show great impact in developing countries. This has especially been shown in health care related cases
 \cite{boulos2011smartphones}, e.g., by providing the opportunity for remote diagnosis based on photos \cite{martinez2008simple}.

\emph{Crowdsourcing} is an increasingly popular way to collect data provided from people at the local level and build larger information bases. The Web 2.0 based platform Ushahidi is one of the most popular examples of the impact of crowdsourcing crisis information \cite{okolloh2009ushahidi}.
In the context of natural disasters, crowdsourcing techniques were used in the 2012 Haiti earthquake to organize help \cite{zook:vgihaiti}. Especially crowdsourcing of geographical information (Volunteered Geographic Information -- VGI) can have a strong impact in developing countries, for example for  monitoring development \cite{mapit-acmdev}.
The impact of VGI was also explored in the case of natural disasters  \cite{Farthing:preparedness10}. The use of VGI in disaster cases is still subject to further extended research \cite{goodchild:vgidisaster10}.

\section{MOBILE4D SYSTEM OVERVIEW}

\label{design}
In this section, we introduce Mobile4D, a mobile crowdsourcing disaster alerting and reporting system implemented in Laos. Mobile4D brings together the power of local knowledge about, e.g., places, people, livestock, or crop with GAUs responsible for coordination and support. Mobile4D enables affected people to report disasters directly to GAUs and enables GAUs to use  direct communication channels to coordinate action and advice. 

\subsection{Situation and information flow in Lao PDR}

The Mobile4D system was planned in tight cooperation with the Ministry of Agriculture and Forestry (MAF) in Lao PDR. It is designed to be embedded in PRAM KSN, a WWW-based knowledge sharing platform among agricultural extension workers \cite{chew:pramksn2013}. In PRAM KSN, extension workers can report on their work, ask questions (that will be answered by teachers and experts in administration), access tutoring material, and, most importantly, get in touch with each other and share experiences and advice directly. Within PRAM KSN, a disaster warning and reporting component was the first enhancement wished by local users of PRAM KSN, as natural hazards play an important role in the extension worker's work in the villages. Mobile4D then was designed to meet the goals and workflows of PRAM KSN.

In a disaster cases, several GAUs in Lao PDR are in charge. For Mobile4D, we concentrate on the administrative units under the Ministry of Agriculture and Forestry (MAF). In Lao PDR, there are 17 provinces, with each province containing a number of districts. MAF has administrative offices in any province (Provincial Agriculture and Forestry Office -- PAFO) and every district (District Agriculture and Forestry Office -- DAFO). All institutions have their specific role in disaster cases. Villages are organized in village clusters, so-called \emph{kumbans}. 

Also, International Non Government Organizations (INGOs) can play a role in disaster cases. Usually, the communication between DAFO and PAFO is pursued by paper. Telephone and fax are only used in urgent cases. Communication between PAFO and Ministry is usually pursued by telephone or fax. Email or other internet services are usually not used, although every province capital has internet access.

However, 3G or 2G mobile internet is accessible in large parts of Laos provided by several telephone providers. With mobile internet connections it is possible to even reach remote locations by TCP/IP services.

\begin{figure*}[t]
\centering
\includegraphics[width=.9\columnwidth]{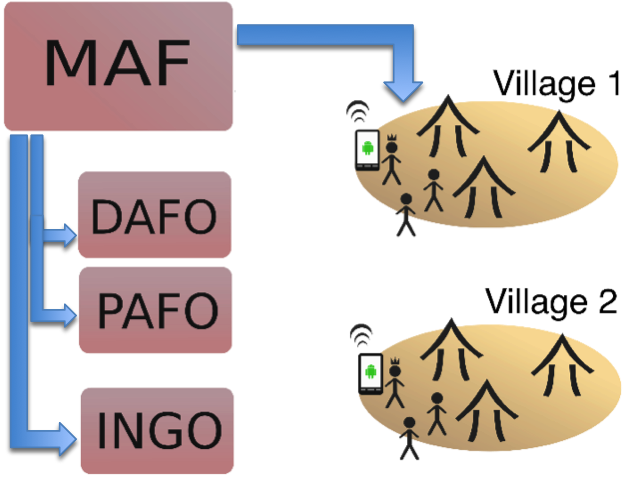}
\hspace{12mm}
\includegraphics[width=.9\columnwidth]{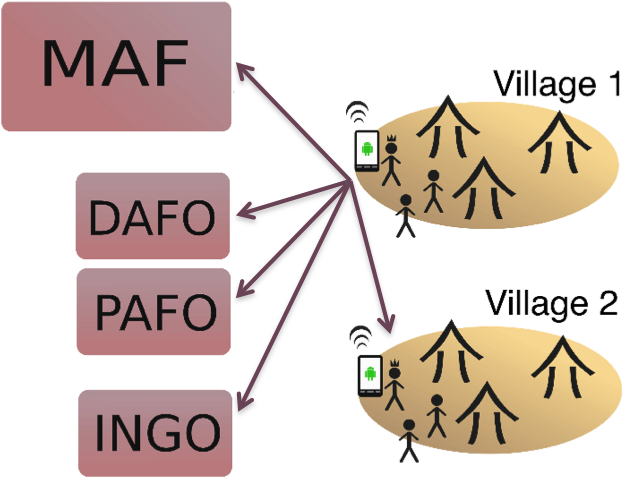}
\caption{Top-down and bottom-up information flow in Mobile4D: information exchange and communication is flexible and can be initiated from GAUs as well from individuals at the village level. In particular, Mobile4D also spreads disaster reports at the local level to other affected users without the need to pass all administrative layers.}
\label{topupbottomdown}
\end{figure*}

\subsection{Main Goals}

ICT systems for disaster management, alerting, and response help, among other things, to overcome shortcomings in communication and information flow. However, those systems (but also any ICT system in general) often neglect the specific affordances of developing countries. Under conditions of limited resources, workflows and systems have to be adapted to local circumstances and cultural background. Thus, the Mobile4D system partiularily aims at two goals:
\begin{enumerate}
\item to provide a \emph{bi-directional} flow of information, that is, both from administration to the local level and vice-versa
\item to institutionalize the flow of communication, that is, to directly embed the reporting and alerting system into administrative workflows.
\end{enumerate}
These two aims tackle problems that have been identified as being amongst the larger challenges for current crowdsourcing based disaster management systems  \cite{narvaez:masterthesis:2012}. Furthermore, Mobile4D aims at
 \begin{enumerate}
\setcounter{enumi}{2}
\item fully integrating small-scale disasters into the usual disaster management workflow by allowing local disaster reports.
\end{enumerate}

\label{overview}
\subsection{System Architecture}

Taking Lao PDR's widespread mobile internet coverage into account, Mobile4D is designed as an internet and smartphone based system for the free and widespread Android platform. The costs for Android smartphones drop drastically, and even older devices offer the full range of sensors and interaction possibilities necessary, making the used market an attractive source for mobile devices.

Mobile4D basically consists of three components:
\begin{enumerate}
\item an \emph{Android app} which allows people in the villages to report disasters, receive warnings, and make contact with people in GAUs to get help,
\item a \emph{WWW frontend} which allows  different GAUs to receive and manage reports, send out warnings and information material, and contact people affected, and
\item the \emph{disaster management server} handling the communication traffic. 
\end{enumerate}

The design of Mobile4D is completely devoted to low-cost technology, realtime communication, and reliability.  The web client runs on several years old PCs with up-to-date web browsers. The web client renders functionality with JavaScript, i.e., all code is loaded from the internet when the page is first accessed. This results in some waiting time at initial startup, but after that, no other network communication is needed than transmitting the efficiently encoded disaster data. This ensures full functionality also under weak network conditions, which is  an issue, because especially district offices rely on mobile internet connections.

First, it is planned to equip the people responsible for disaster response at the district level with a smartphone, as these are the people that take over disaster reporting duties anyway (that is, one smartphone for every district would suffice). But in a longer perspective and with increased use of smartphones, every individual with an Android smartphone is a potential contributor to the Mobile4D system and is also able to receive disaster warnings.


\subsection{Information Flow and Sharing}

Mobile4D 
supports direct top-down and bottom-up communication to exchange information. Information can be disaster alerts, information material, media, or ongoing correspondence about situations between GAUs and people affected. Figure~\ref{topupbottomdown} shows how information can flow \emph{top-down} from the ministry level (MAF) directly to specific affected villages and back. Additionally, the information is distributed to the correct subordinate units on province (PAFO) and district (DAFO) level, and to non-governmental organizations (INGO). These direct channels enable to shortcut slow information distribution and make information available immediately where it is required. All alerts are sent out as Push messages. This ensures  reliable real-time communication while being extremely efficient in terms of bandwidth.

\begin{figure}
\centering
\includegraphics[width=.48\columnwidth]{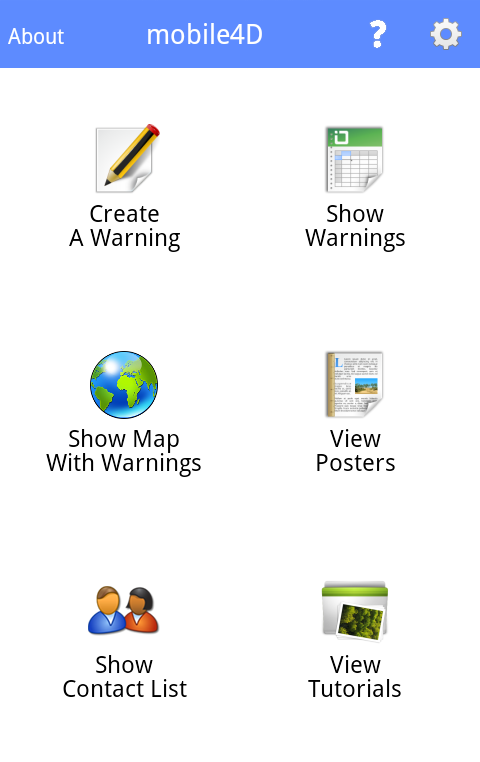}
\hspace{1mm}
\includegraphics[width=.48\columnwidth]{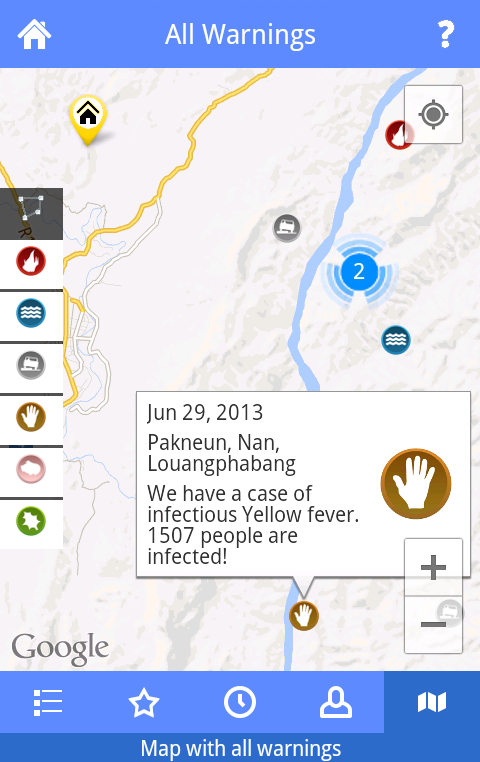}
\caption{Screens of the Mobile 4D Android app. Left: Start screen with direct access to all important functions. Right: Map overview of nearby disaster alerts.}
\label{screenshots-android}
\end{figure}

Most importantly, our crowdsourcing approach enables information to flow in the same way \emph{bottom-up}: when people are affected by disasters, they can report on the situation to the GAUs and INGOs. This is supposed to be done by the Android app which allows to send reports directly from the place where the disaster has occurred (see Fig.~\ref{screenshots-android} for screenshots). Mobile4D sends the reports to all the responsible GAUs in the hierarchy, but directs the information to the GAU responsible to take action (e.g., infrastructural problems can be resolved on district level, while severe disease outbreaks are handled on province level). Internal protocols ensure that information gets reviewed and answered. Also reporters will always be automatically notified when their report is processed by staff in the GAU. Whenever a disaster is reported at local level, the information about it is immediately sent out to all neighboring villages without administrative review. People get informed when situations are reported and can guard against potential threads to protect health and belongings at a very early stage.  

All information can be shared with everybody by forwarding received information via SMS or establishing a voice call. Phone numbers of local reporters and GAU staff are always prominently displayed and can be used for direct communication by everyone. Furthermore, Mobile4D supports interfacing with social network platforms like Twitter, where detailed information about disasters and their states can be made available along with their geographic location. 

\subsection{Information Generation and Processing}
\begin{figure}[t]
\centering
\includegraphics[width=.48\columnwidth]{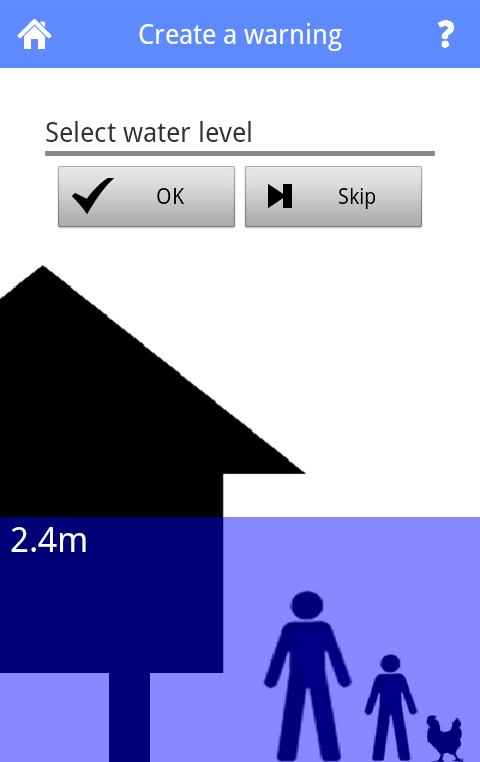}
\caption{Intuitive text-free interface for entering the water level in a flooding by sliding the water up and down.}
\label{screenshots-water}
\end{figure}

In contrast to social network based platforms, Mobile4D has a particular focus on the data gathering process. It offers structured assembly of information to create specific disaster alerts (floods, bush fires, infrastructural problems, and diseases of humans, animals, and plants). In both, the mobile and the web client users are guided stepwise through an intuitive disaster reporting process.  This step-by-step procedure ensures that important data is at least asked for and helps the reporters to provide structured information even in stressful situations.
As it has been shown that  text-free widgets can usually be fully understood \cite{MUIdesign}, Mobile4D tries to avoid textual interfaces wherever possible (see Fig. \ref{screenshots-water} for an example).


\subsection{Administrative Integration}

\begin{figure*}
\centering
\includegraphics[width=2\columnwidth]{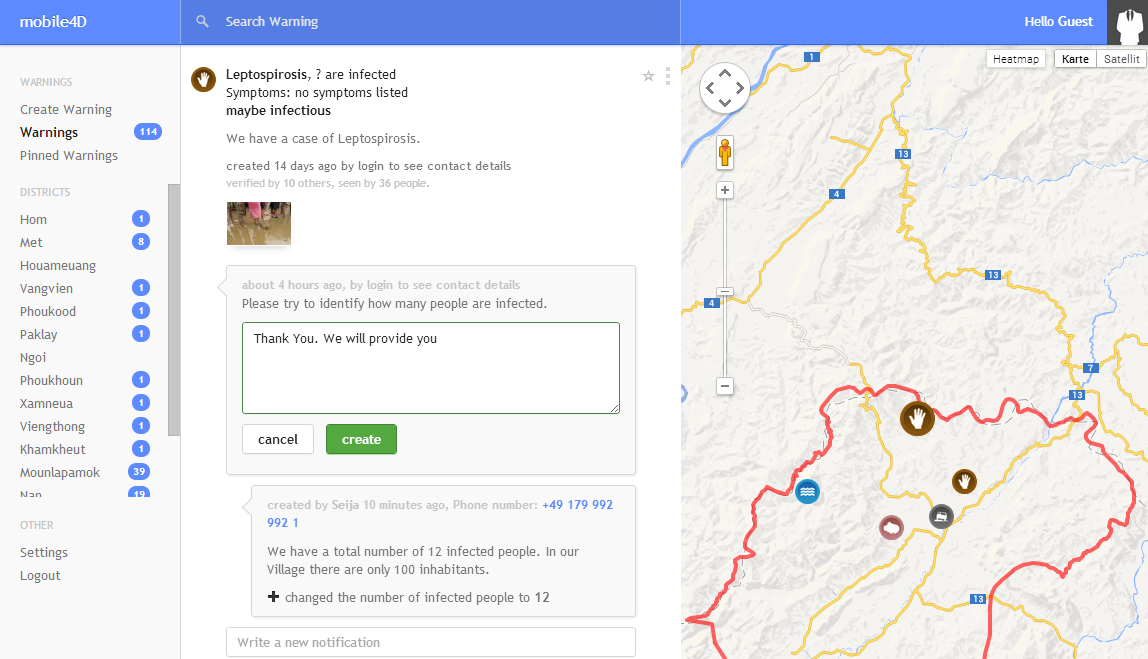}
\caption{Screenshot of the administrative WWW interface. It opens a communication channel that allows to ask further questions or send information to the reporter (and anyone affected), but also offers contact information like telephone numbers for direct contact. An overview of reported disaster cases is shown as icons on a map.}
\label{www}
\end{figure*}

Staff of GAUs have a role-based web browser interface (see Fig. \ref{www}) to process incoming information of local reporters or other GAUs. They can establish direct communication, send out information and support to places and people in their area of responsibility. They are able to add tutorial documents (e.g., in PDF format) to disaster reports that will directly be spread to the smartphones of affected users. Mobile4D provides tools to perform any kind of administrative work related to disasters: reviewing information, getting in touch with reporters, assigning issues to other administrative layers, sending out information material, updating, merging, resolving disasters, etc..

At the current level, Mobile4D does not distinguish different administrative layers more than providing a specific role. As we believe in the power of local solutions, each GAU is a fully autonomous participant within the system. All layers can access the same data. It is consequently monitored which GAU performed which action, and all other GAU will be notified if they feel responsible for the disaster edited. That being said, Mobile4D does not enforce new workflows within the administration, but if offers full transparency for all administrative layers to inspect and monitor actions taken by others. In any step, contact information is provided, and the possibility to get in direct contact, through the system or any other communication channel, is encouraged.

To assure the quality of data, Mobile4D provides a multi-leveled verification system based on the different administrative layers: Each GAU (MAF, PAFO, DAFO) has the opportunity to \emph{verify} any report, e.g., after checking back via a phone call or personal visit, thus giving the report and "official" stamp. In addition, single users can also verify the report. That is, the crowd is used for quality assurance itself, as a large number of user verification is a good estimate of its reliability \cite{freifeld2010participatory}. For the user, this verification system can be a valuable help to assess the reliability of data on his own. Perspectively, this verification mechanism can also be used by the system to automatically decide when an alert is distributed without any administrative interference. At the moment, this is not implemented yet.

\subsection{Interoperability}

\subsubsection{Compliance with the Common Alerting Protocol}

Mobile4D does not aim at providing all the features a full-fledged disaster management system can offer. Especially in the response phase, powerful management systems are available. To be able to be interoperable with such systems, Mobile4D is fully compliant with the Common Alerting Protocol (CAP) \cite{jones2005common}. CAP is an XML based protocol for exchanging warnings and alerts. CAP has become an OASIS\footnote{OASIS: Organization for the Advancement of Structured Information Standards} standard in 2004.

Mobile4D fully maps a disaster's attributes (such as sender, urgency, status) to the corresponding fields in CAP. Attributes  supported in Mobile4D only but not in CAP (such as information specific to the location, for example, the name of the village cluster (kumban) where the disaster occured) are stored in the CAP {\tt parameter} field not to lose important information. By that, Mobile4D allows for export of any alert into CAP, and alerts specified in CAP can be imported into the CAP database. It is easily possible to adapt the Mobile4D API to automatically accept CAP specified alerts.

The following code sniplet shows the result of a Mobile4D disaster report exported in CAP:
\lstset{language=XML}
\begin{lstlisting}
<?xml version="1.0" encoding="UTF-8"?>
<alert xmlns="urn:oasis:names:tc:emergency:cap:1.1">
   <identifier>7</identifier>
   <sender>89</sender>
   <sent>2013-09-25T07:05:02.917-05:00</sent>
   <status>Actual</status>
   <msgType>Alert</msgType>
   <source>MAF office; +856 1234567; MAF</source>
   <scope>Public</scope>
   <info>
      <language>en-US</language>
      <category>Health</category>
      <event>I have seen the same thing in another 
             village nearby last year</event>
      <responseType>None</responseType>
      <urgency>Future</urgency>
      <severity>Extreme</severity>
      <certainty>Possible</certainty>
      <effective>2013-09-24T19:00:00-05:00
      </effective>
      <parameter>
         <valueName>location</valueName>
         <value>19.845519,102.078652</value>
      </parameter>
      <parameter>
         <valueName>disasterType</valueName>
         <value>PlantDiseaseInfo</value>
      </parameter>
      <parameter>
         <valueName>province</valueName>
         <value>Louangphabang</value>
      </parameter>
      <parameter>
         <valueName>district</valueName>
         <value>Louangprabang</value>
      </parameter>
      <parameter>
         <valueName>kumban</valueName>
         <value>Sangkalok</value>
      </parameter>
   </info>
</alert>
\end{lstlisting}

\subsubsection{Geocoding with MapIT}
Mobile4D allows to geocode pictures with MapIT \cite{mapit-agile}. MapIT is a tool to generate geometric geographic data directly from photos taken with smartphones. This feature is helpful to locate, e.g., agricultural lots with their precise geometry.  MapIT allows to directly mark the object of interest on the smartphone picture, and the resulting geographical object is directly integrated in  Mobile4D disaster alerts as the area being affected. This allows for very exact localization of  alerts.

\begin{figure*}
\centering
\includegraphics[width=.68\columnwidth]{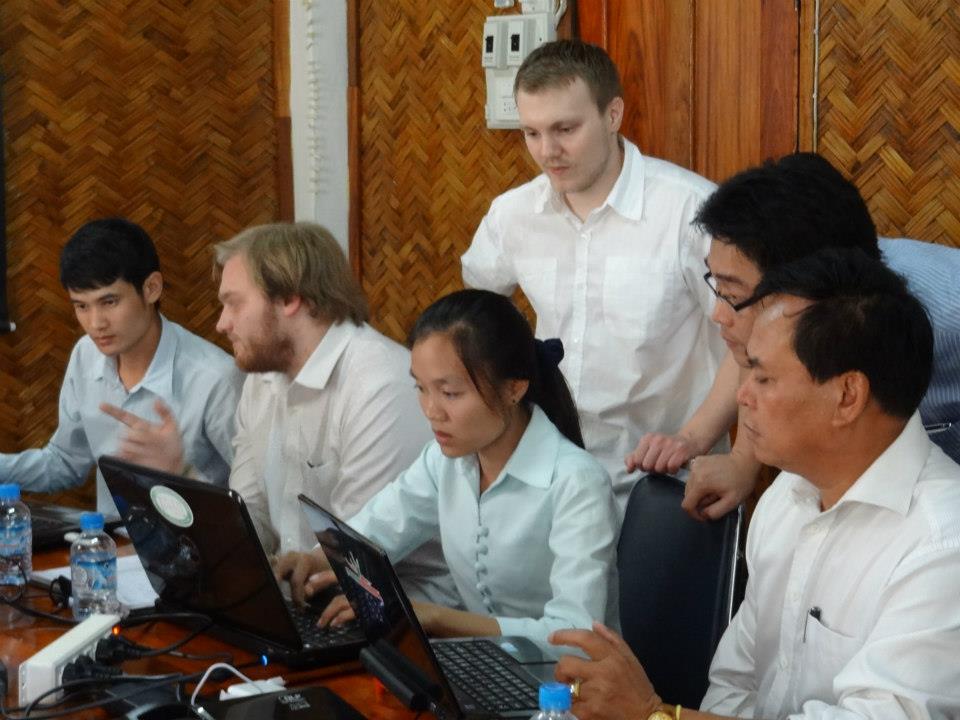}
\includegraphics[width=.68\columnwidth]{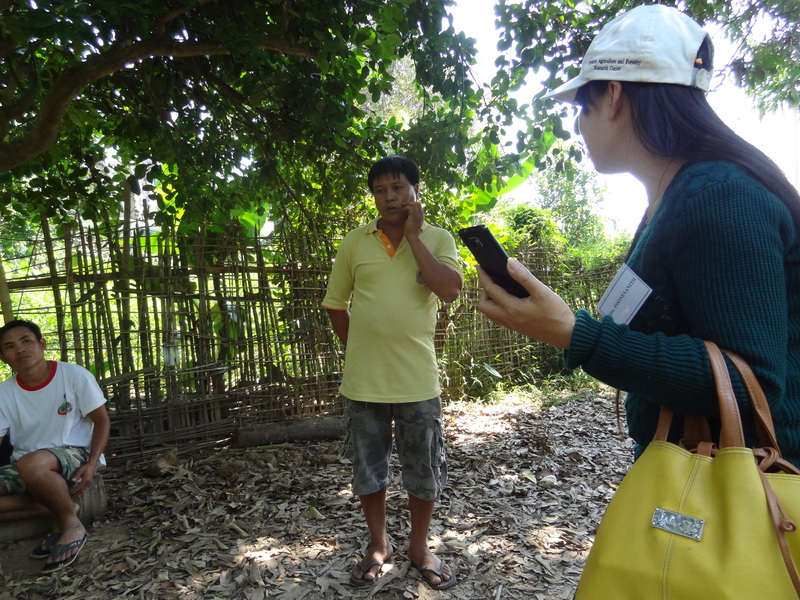}
\vspace{1mm} \\
\includegraphics[width=.68\columnwidth]{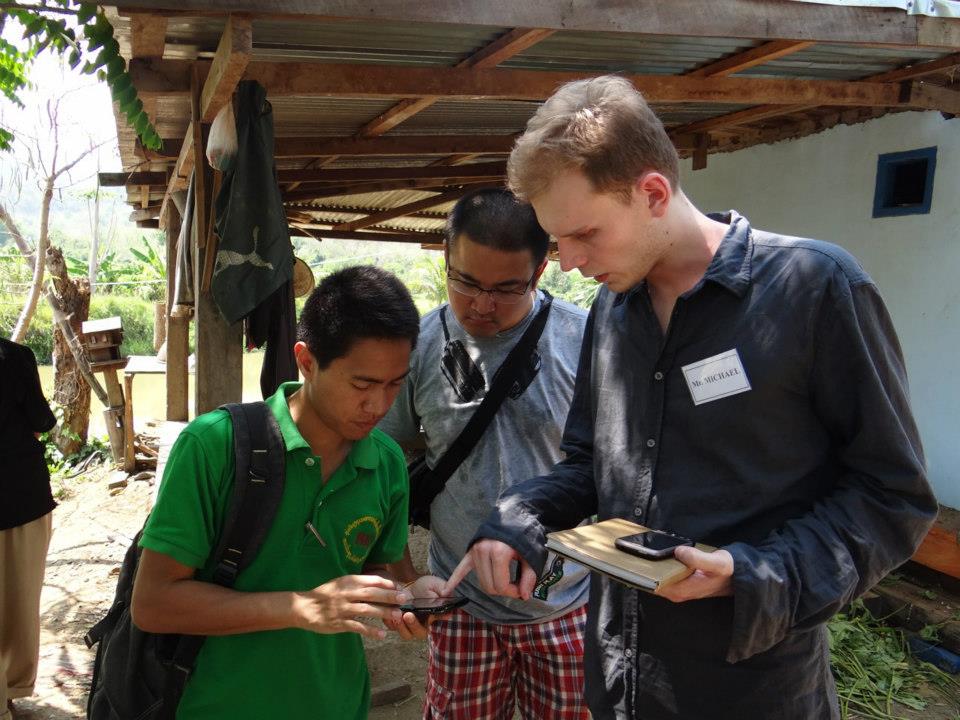}
\includegraphics[width=.68\columnwidth]{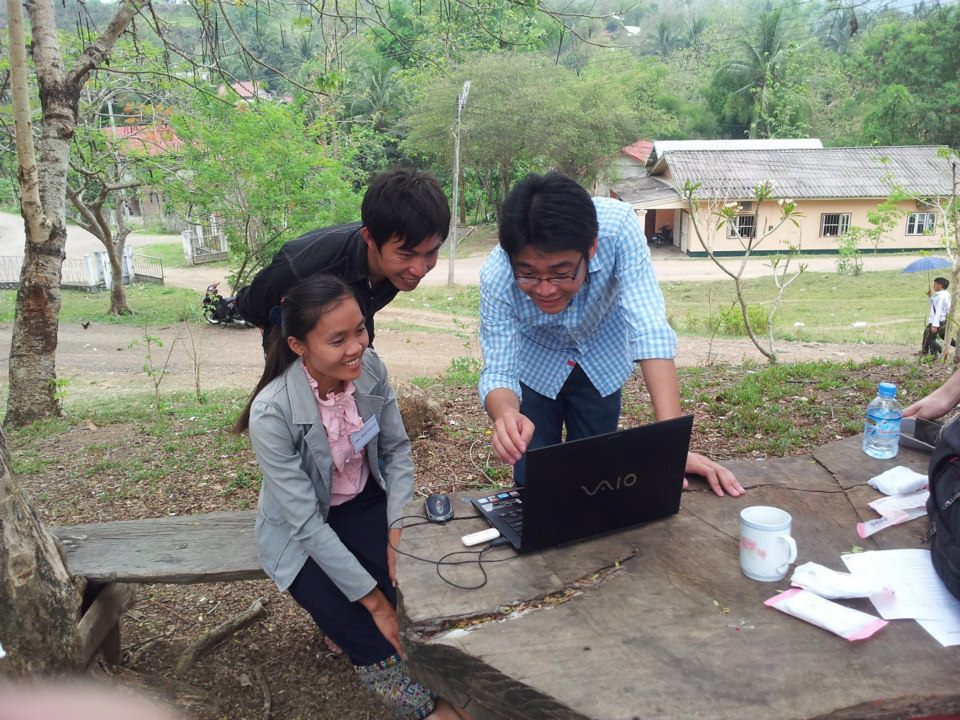}
\caption{Mobile4D field tests in Laos: training and data acquisition in the districts Luang Prabang, Chompet und Pak-Ou.}
\label{fieldtest}
\end{figure*}

\section{MOBILE4D TESTS IN LAO PDR}
\label{results}
Mobile4D was extensively tested in April 2013  in the province of Luang Prabang, Lao PDR (see Figure \ref{fieldtest}). The test involved staff members of MAF, staff members of the province office of Luang Prabang, and district officers of districts in the province. The system was set up with locally available technical infrastructure, that is, laptops used at work and privately owned smartphones. We used mobile internet provided by three different phone companies. This resulted in a highly heterogeneous technical ecosystem. Main purpose of the tests was to gather feedback from the people directly affected with disaster alerting and management at the administrative levels, as they are the prospective users of the system. The system proved to work reliably and efficiently, also under very weak network conditions.

During extensive feedback sessions with all participants we identified points of improvement. Most prominently, those were in the area of information visualization and usability. Further features  demanded were integrating more reasoning and forecasting capabilities and the possibility to edit and add geographical information such as place names. All participants were very positive about the Mobile4D system and hoped for being able to use in as a part of their daily routines. In particular, they pointed out the efficiency of direct communication channels between affected people and GAUs, which allows to take quick actions and provide important information directly where it is needed. As a result of the successful tests, it was agreed with MAF to set up a pilot installation in one province in Laos to evaluate the system's impact for a longer period of time.

\section{CONCLUSIONS}

We presented Mobile4D, a crowdsourced system for disaster reporting and alerting with smartphones. Along with larger natural disasters, Mobile4D targets also at small-scale hazards at a local level. It allows for bi-directional communication, from the local level towards administration and vice-versa. Governmental administrative units are directly involved in the flow of data, and local communication structures are strengthened. In an extended field test, Mobile4D performed reliably and efficiently, proving its suitability for developing countries such as Lao PDR.

\addtolength{\textheight}{-14.6cm}   





\section*{ACKNOWLEDGMENT}

Part of this work is supported by the German Research Foundation (DFG) through the Collaborate Research Center SFB/TR\,8 "Spatial Cognition". Further funding was provided by the German Ministry for Research and Education (BMBF). We thank the Lao Ministry for Agriculture and Forestry for substantial support, in particular Savanh Hanephom, Thatheva Saphangthong, Soudchay Nhouyvanisvong, and Alounxay Onta, as well as  our partners at UNU-IIST Macau, Peter Haddawy, Han Ei Chew, and Borort Sort. We also want to give credit to the students of the Mobile4D student project at the University of Bremen for their dedicated work on the system: Timo Bonanaty, Christian Czotscher, Nathalie Gabor, Satia Herfert, Helmar Hutschenreuter, Andreas K\"astner, Pascal Kn\"uppel, Daniel Langerenken, Carsten Pfeffer, Thorben Schiller, Arne Schlamann, Urs-Bj\"orn Schmidt, Nadine Schomaker, Denis Szadkowski, Denny Teuchert, Thomas Weber, Malte Wellmann, Michal Wladysiak, and Daniela Zimmermann.


\bibliographystyle{abbrv}
\bibliography{literature/disaster,literature/mapit}

\end{document}